# A Novel Hybrid Fast Switching Adaptive No Delay Tanlock Loop Frequency Synthesizer

Ehab Salahat, Saleh R. Al-Araji, Mahmoud Al-Qutayri
Department of Electrical and Computer Engineering, Khalifa University, Abu Dhabi, U.A.E.
ehab.salahat@ieee.org

*Abstract*—**This paper presents a new fast switching hybrid frequency synthesizer with wide locking range. The hybrid synthesizer is based on the tanlock loop with no delay block (NDTL) and is capable of integer as well as fractional frequency division. The system maintains the in-lock state following the division process using an efficient adaptation mechanism. The fast switching and acquisition as well as the wide locking range and the robust jitter performance of the new hybrid NDTL synthesizer outperforms conventional time-delay tanlock loop (TDTL) synthesizer by orders of magnitude, making it attractive for synthesis even in Doppler environment. The performance of the hybrid synthesizer was evaluated under various conditions and the results demonstrate that it achieves the desired frequency division.**

## I. INTRODUCTION

**FREQUENCY** synthesizers, both integer and fractional types, are fundamental building components in many contemporary wireless communication systems, clock generators, and a multitude of disparate applications. Data transfer involves modulation at the transmitter side on an RF carrier, and demodulation at the receiver, therefore an accurate RF carrier must be generated and a synthesizer is essential for channel selection and frequency translation [1] [2] [3]. The architecture and the design of synthesizers present major challenges due to the stringent RF requirements as well as the demands for high speed in digital transceivers supporting the drive for high resolution, wide bandwidth and fast switching speed. Conventional synthesis techniques are categorized into three classes: Direct Analog Synthesis (DAS), Direct Digital Synthesis (DDS), and Phase-locked loop (PLL) based, or Indirect Synthesis [1]. Each of these types has its merits and drawbacks. PLL based frequency synthesis has been widely used in industry. The main functional components of this type are the Phase Detector (PD), the Voltage Controlled Oscillator (VCO), the Loop Filter (LF), and the Frequency Divider [4]. However, a drawback of the PLL-based technique is that a PLL with a wide frequency range is difficult to achieve. With fractional-*N* synthesis technique, finer frequency control can be achieved; however, these systems typically have very narrow bandwidth [5] [6] [7] [8]. Depending on the sampling process used, digital phase lock loops (DPLLs) are classified as uniform or non-uniform [1]. Non-uniform techniques tend to be more attractive as they ease the system modeling and subsequent circuit implementation of DPLL with an offer of better acquisition time compared to the uniform ones [9]. In [10], a new architecture of an all-digital fractional-N PLL frequency synthesizer was presented, employing an extra time-to-digital converter (TDC) to measure the fractional value. In [11], an alternative approach uses a delta-sigma frequency-to-digital converter (ΔΣ-FDC) in place of a TDC to retain the benefits of TDC-PLLs and ΔΣ-PLLs, whereas the authors in [12] explore a new topology of charge-pump PLL intended for ΔΣ-fractional-*N* frequency synthesis. A new scheme in [13] is proposed to produce spectrally pure clock signal at certain frequencies, utilizing the so called Flying-Adder synthesizer as a fractional divider placed inside the PLL loop. Moreover, the authors in [14] has also proposed a new Zero-Crossing DPLL (ZC-DPLL) frequency synthesizer, with an excellent performance and low-complexity, which can also adopt the hyperbolic nonlinearity proposed in [15] as well.

The Time Delay Digital Tanlock Loop (TDTL) proposed initially in [16] is a non-uniform type of Digital PLL and is shown in Fig. 1. The use of the TDTL for frequency synthesis was demonstrated in [17]. In that work, the authors proposed using adaptive gain and register-based mechanisms to have the TDTL perform frequency synthesis. The register-based technique achieved better results. The TDTL architecture in Fig. 1 includes a time delay block. This block introduces nonlinearity to the lock range characteristic of the loop. This nonlinearity has been overcome by the NDTL (No Delay Tanlock Loop) architecture proposed in [1] [18].

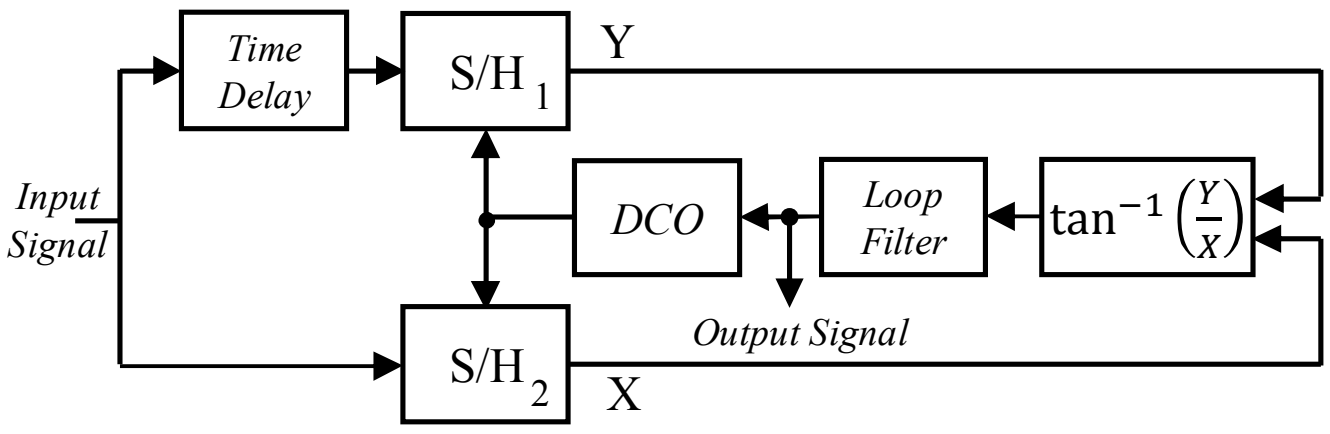

Fig. 1: Time Delay Tanlock Loop (TDTL) Architecture **[1]**.

In this work, a new frequency synthesizer is proposed based on the NDTL architecture that can perform both integer and fractional synthesis using the same frequency division mechanism. The proposed NDTL synthesizer has no nonlinearity, and is characterized by its fast acquisition and wide locking range.

The remaining part of the paper will be presented as follows: Section II presents the architecture and mathematical analysis of the proposed NDTL synthesizer. Section III shows the analysis of the simulations results. Finally, the paper findings are summarized in section IV.

## II. SYSTEM ANALYSIS AND ARCHETICTURE

### *A. NDTL Archeticture Overview*

The architecture of the digital tanlock loop with no delay NDTL is shown in Fig. 2. This architecture consists of two sampling blocks, loop filter, an arctan phase detector, and a modified version of the DCO with two sampling signals that have a quadrature relationship. The center frequency of the DCO, which is part of the loop DCO block, is set at twice the overall free-running frequency ($f_o$). The DCO output signal is used to drive two counters that are triggered at positive and negative edges respectively, maintaining the quadrature relationship and eliminating the need for the time delay block. The outputs of those counters are used to sample the input signal. Detailed analysis of the NDTL architecture is provided in [18]. Fig. 3 illustrates the locking range of the NDTL and compares it with that of the first order TDTL.

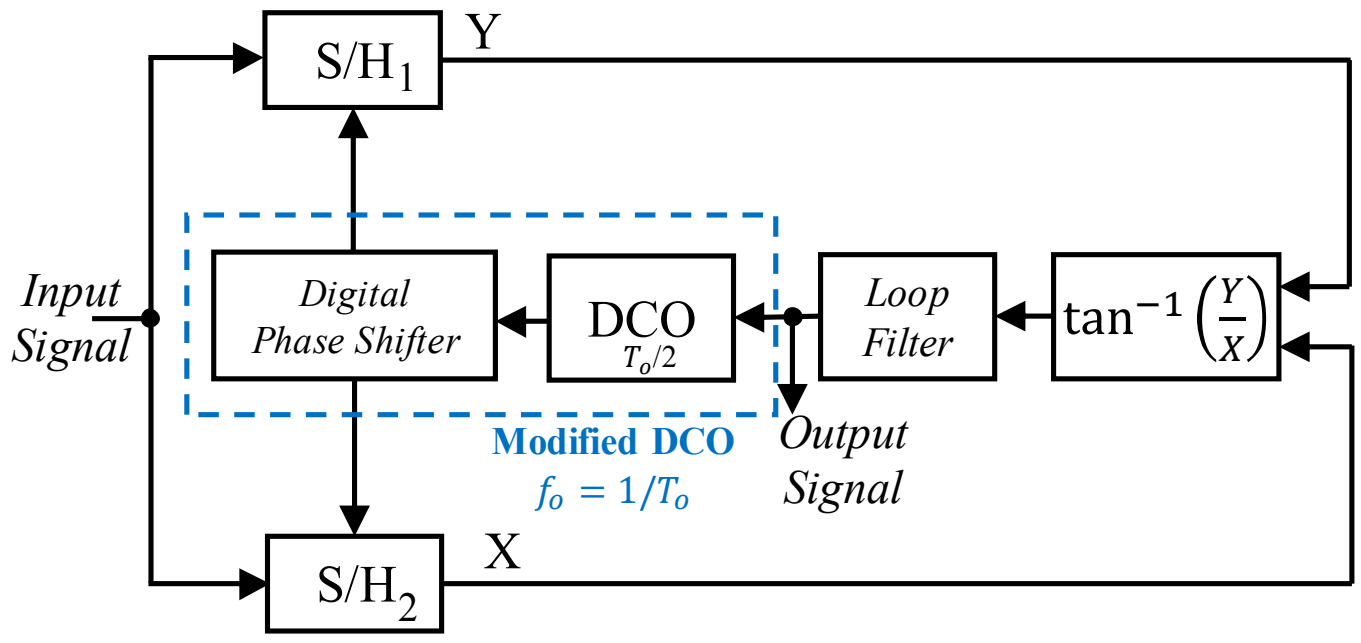

Fig. 2: No Delay Tanlock Loop Architecture [1].

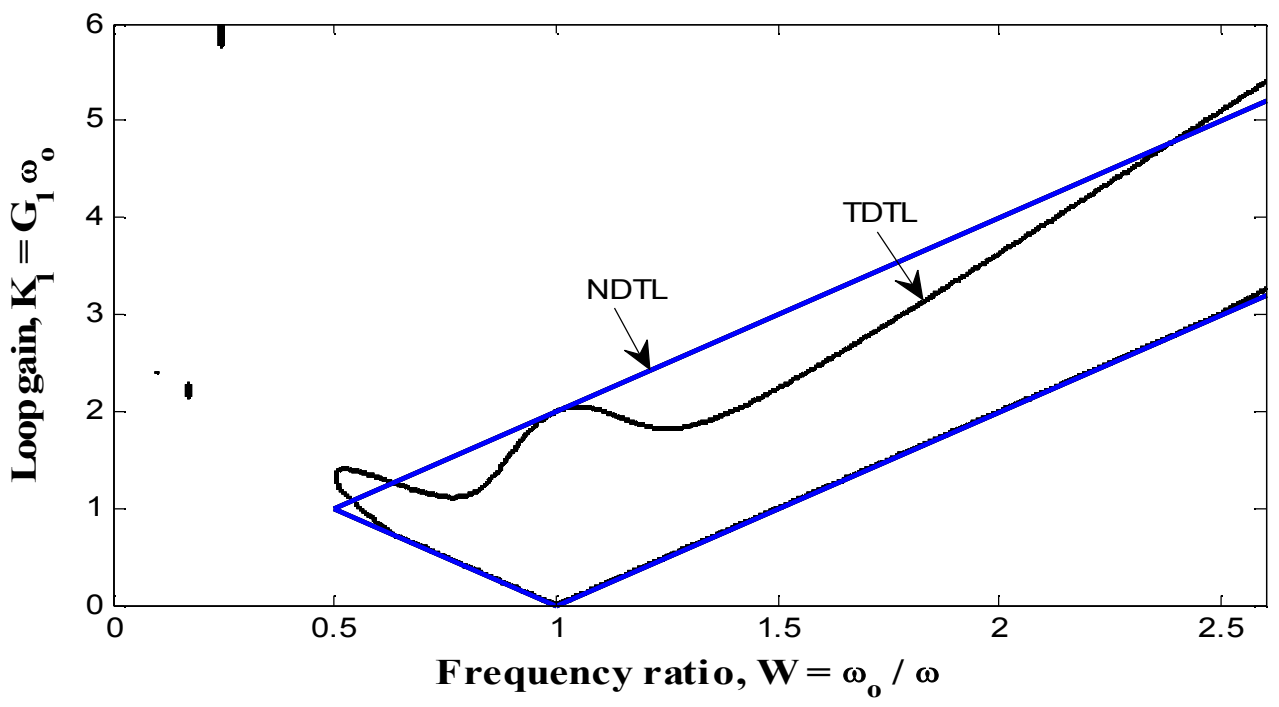

Fig. 3: Locking Range comparison between TDTL and NDTL [1].

### *B. Hybrid NDTL Frequency Synthesizer Architecture*

The architecture of the Hybrid NDTL Frequency Synthesizer is shown in Fig. 4. This architecture implements an indirect frequency synthesis. The main block in this architecture is the divider block, which is a counter. Those divider blocks are placed between the DCO block and the Sample and Hold (S/H) blocks. The DCO output signal derives two divider blocks that are configured exactly the same (Integer or Fractional), except that the upper one is negative edge triggered whereas the lower one is a positive edge triggered. Hence, the output frequency of the divider block is a multiple of the DCO frequency. This output frequency of the divider block is now responsible for the sampling process, rather than the DCO frequency. As in the case of the NDTL design, the positive and negative edge triggering is essential to maintain quadrature relation between the two arms of the loop, since a delay block is not used anymore.

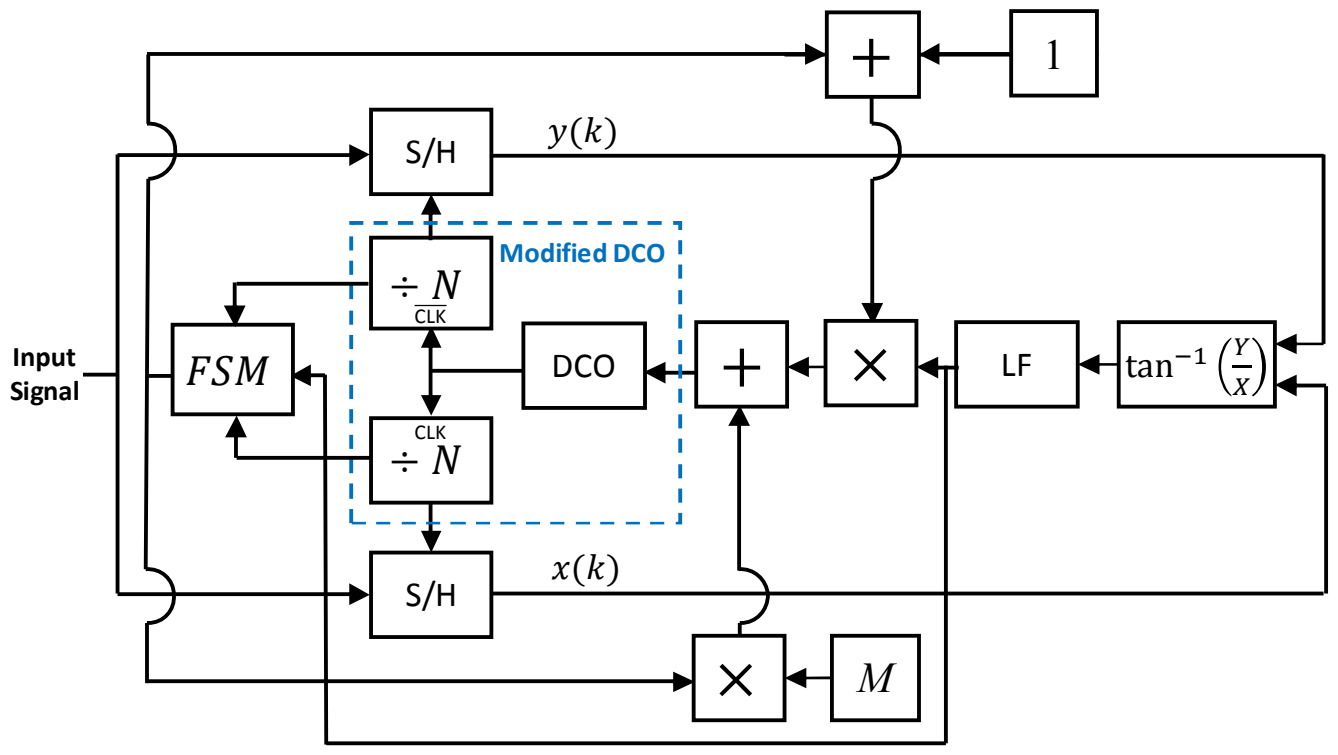

Fig. 4: Hybrid NDTL Frequency Synthesizers Architecture [1].

As stated earlier, the divider is hybrid (i.e. can be integer or fractional). The division process is effectively a counting process. For integer frequency synthesis, the divider produces only one output frequency relative to the counting value set in the counter. When the counting value is set to *N,* which means there are *N*+1 states including the zero, the output frequency will be the DCO frequency divided by this *N*+1*,* which is the operation of integer frequency synthesis. For example, when the frequency desired at the output of the divider has 1:3 relation with the DCO frequency, then the value of *N* is set to 2 (counting states: 0, 1 and 2) and so on.

As for the fractional frequency synthesis, the process of division becomes somewhat more involved. Since counters produce only integer values and no fractions, three counters will be used to produce, on average, the desired fraction that will be used as a division factor, keeping in mind that any fraction lies in between two consecutive integer values. When the division factor is set to *N.F* (i.e., *Integer. Fraction*), two consecutive integer values are used, $P_1$ and $P_2 = P_1 + 1$. The fractional divider output will be exclusively one of them but never both. For each fraction there is a pattern of $P_1$ and $P_2$ that produces the desired fraction, which is controlled by the third counter. For example, when the desired division fraction is 1.333, then $P_1 = 1$ and $P_2 = 2$, $P_1$ is passed two thirds the time and $P_2$ in the last third, (i.e., [1 +1 +2]/3=1.3333), and generalization is straightforward. Figure 5 shows the hybrid divider that is used in the architecture of the Hybrid NDTL Synthesizer. For the integer configuration, $P_2$ is set to be exactly as $P_1$ rather than being more by one. The upper counter refers to the $P_1$ integer value, the last counter refers to the $P_2$ integer value and the counter in the middle controls, with the assistance of the other logic circuits, the division process. Note that the hybrid divider has two outputs, HIT which will be used with the S/H block, and the DIV, which is needed to ensure that the operating point is within the locking range as will be described and discussed later.

### *C. Optimum Operating Point of the System*

Referring to Fig. 3, the optimum operating point for the system can be located in the locking range where $W = \omega/\omega_0$ and $K_1 = G_1\omega_0$ are both unity. However, the division process of the dividers may move the operating point to a new location outside the locking range, which causes the system to go out of lock. This is because the DCO frequency, $\omega_0$, has been divided by a factor $\beta = DIV \in \{P_1, P_2\}$, which is an integer value. This means that the new values for $W$ and $K_1$ respectively are:

$$W = \frac{[\omega_0/\beta]}{\omega} = \frac{\omega_0}{\beta\omega} \quad (1)$$

$$K_1 = \frac{G_1\omega_0}{\beta} \quad (2)$$

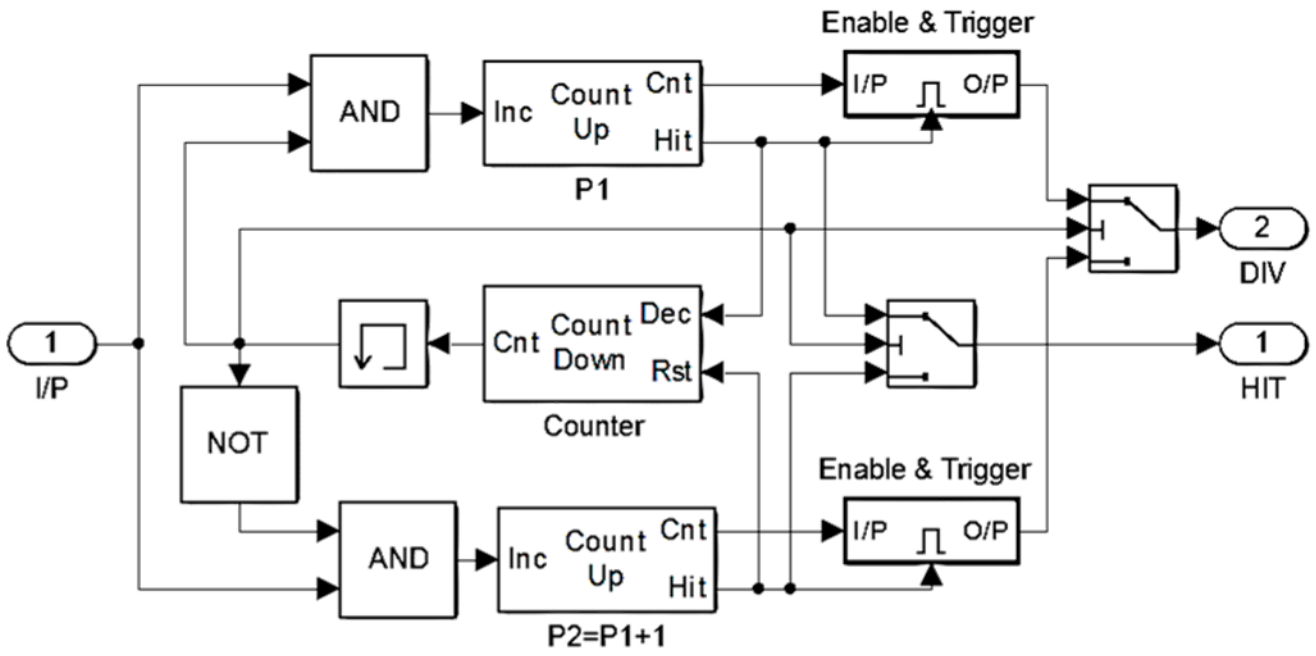

Fig. 5: Hybrid Divider Architecture **[14]**.

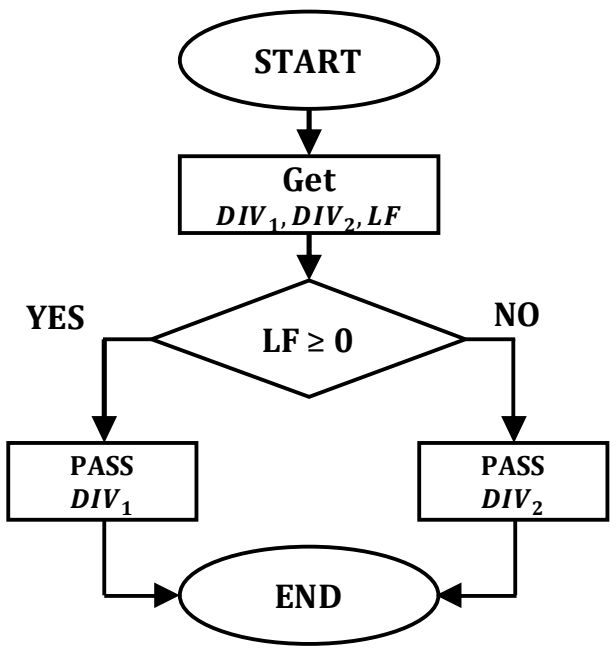

Fig. 6: FSM Flowchart **[14]**.

It can be concluded from (1) and (2) that the optimum operating point has been moved in Fig. 3 by a factor of $\beta$. This is due to the fact that the sampling frequency is no longer $\omega_0$. In order to compensate for $\beta$ in (1) and (2), the gain has to be multiplied by $\beta$, and the DCO frequency has to be increased to reach $\omega_0$ again. Hence, to achieve this for the loop filter, (the gain block in this case), the gain value is simply multiplied by $\beta + 1$, and some $\beta$ DC is added to the input of the DCO to shift the center frequency to be $\beta\omega_0$. This effectively restores the optimum operating point of the system. Note that the DC increase at the input of the DCO block can be achieved simply by making the value of the constant *M* in Fig. 4 to be 1 and use the sensitivity of the DCO equal to the center frequency of the DCO, or a compromise between the value of *M* and the *DCO* sensitivity, which gives an additional option to the designer and allows more design flexibility.

### *D. The Finite State Machine*

The purpose of the FSM is to enhance the acquisition speed and the overall system performance. This is an essential requirement especially when the incoming sinusoid was triggered by a negative step value. The flow chart of the FSM algorithm is shown in Fig. 6. The FSM has three inputs, the two DIV values coming from the fractional dividers and the output of the loop filter. When the FSM senses that the deriving input step is positive, it uses the DIV value from the negative edge triggered divider, otherwise it uses the second DIV value.

## III. Simulation Results

The system has been tested in noise-free environment with different values of input signals. Three ways are used to verify the functionality of the system, namely by observing the phase plane, the loop filter output and the output of the dividers relative to the DCO frequency. Configuring the system to achieve integer synthesis and divide by 4 ($P_1 = P_2 = 3$), with an input step of 0.2 volts, figures 7, 8 and 9 are generated. Figure 7 compares the input step with the output of the loop filter. It can be seen from this figure that the system acquires lock quickly and reaches the steady state value. Figure 8 shows the phase plane of the system ensuring the stability of the system, whereas Fig. 9 shows that there's a 1:4 relation between the DCO and output of the dividers' frequency, which proves the system functionality as an integer synthesizer.

Another test scenario uses a step of 0.3 volts and a division factor of 4.2. Figures 10 and 11 are generated with this configuration. It can be seen from Fig. 10 that the system locks under these conditions and reaches its steady state value, whereas Fig. 11 illustrates the process of fractional division, in which the DCO frequency is divided by 4 four times and in the fifth time it gets divided by 5, which, on average is 4.2. One final test will be done using a negative step value (-0.3 volts) with fractional synthesis (÷3.125). Figures 12 and 13 are generated using this configuration. It can be seen that the system provides an excellent performance in terms of fast acquisition and wide locking range. The phase plane in Figure 13 proves the system stability.

Fig. 14 shows the jitter performance comparison between the TDTL and NDTL synthesizer, for input step value of 0.3 volts, VCO frequency of 100 Hz, M=3.125, VCO sensitivity of 32V/Hz, (note that 32×3.125=100) and division factor=4.2. One can clearly see that the jitter performance of the NDTL synthesizer (ranging from $10^{-2}$ to $10^{-5}$ seconds) outperforms that of the TDTL (that ranges $10^{-1}$ to $10^{-3}$ seconds) by orders of magnitude, making this synthesizer an excellent candidate for many jitter-sensitive applications.

## IV. Conclusions

In this paper, a new efficient and fast switching hybrid frequency synthesizer design is presented based on the No Delay Tanlock Loop (NDTL) architecture. The system is capable of achieving robust integer as well as fractional frequency devision. It utlizes an adaptation mechanism with relatively low complexity in order to keep the system in lock as it performs the devision process. Time Jitter study in AWGN environment comfirms system's reliabiltiy, which outperforms TDTL synthesizers by orders of magnitudes, which makes it an excellent candidate for synthesis even in high Doppler environment.

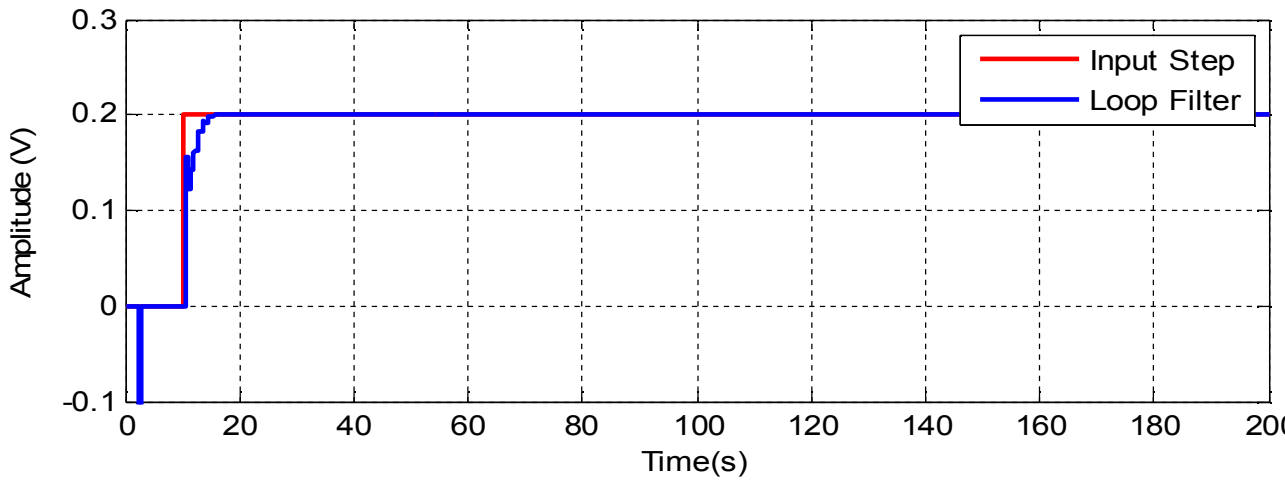

Fig. 7: LF output vs. Input step of 0.2 V, Integer Synthesis (÷4).

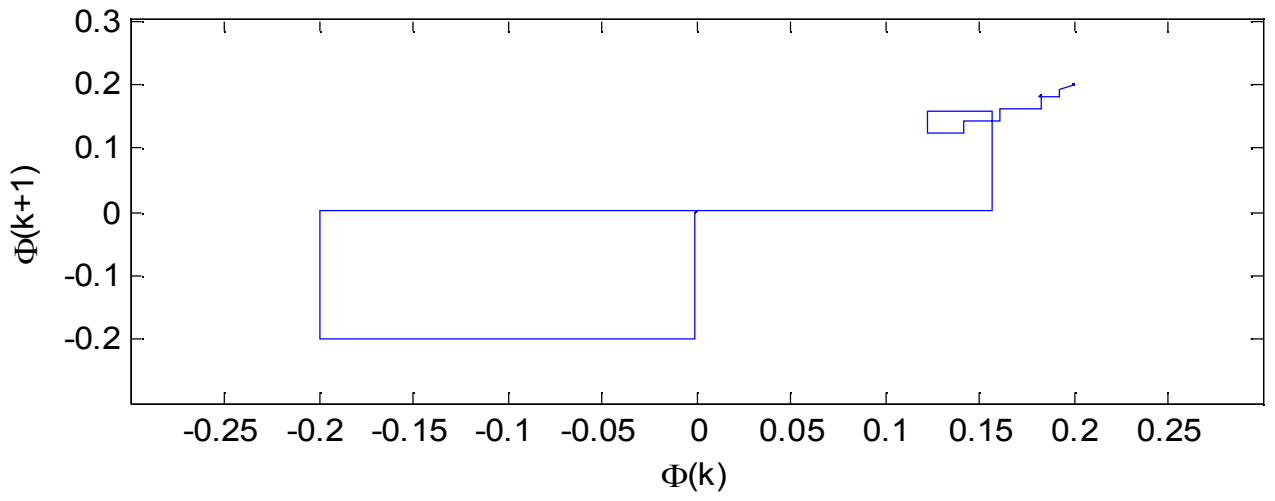

Fig. 8: Phase Plane for step of 0.2 V, ÷4 synthesis.

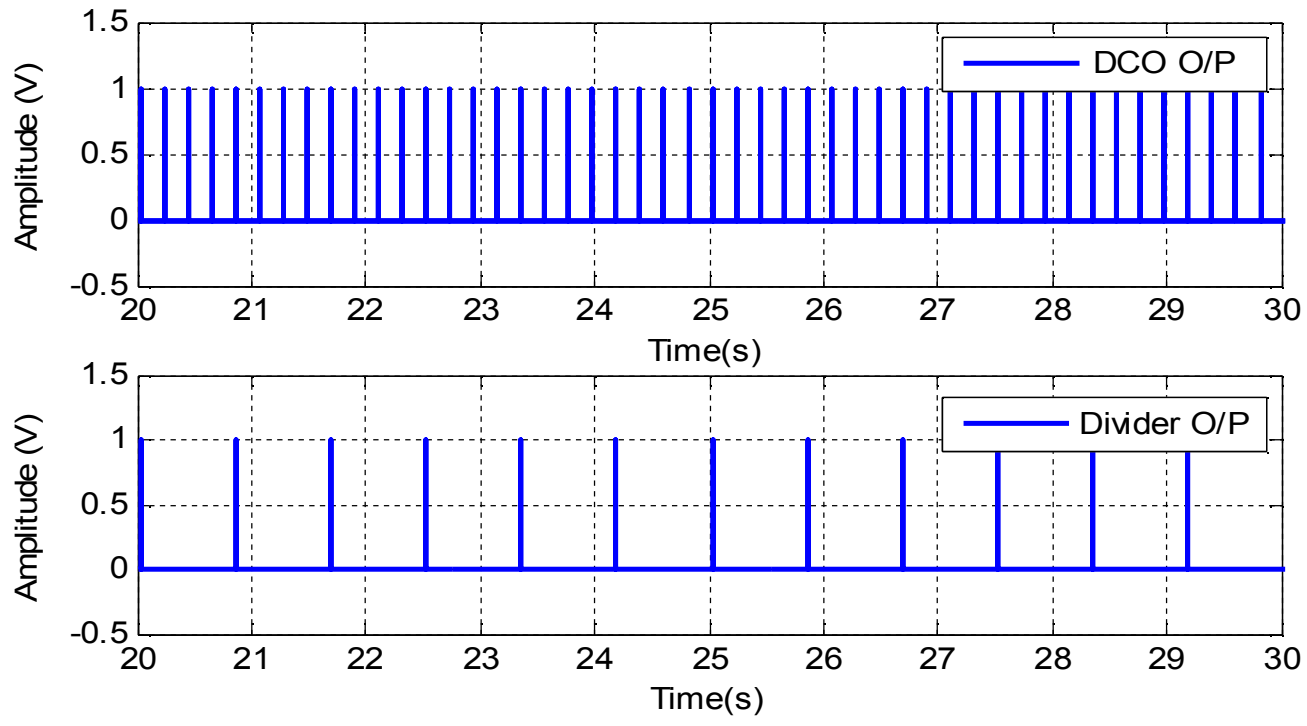

Fig. 9: Divider freq. vs. DCO freq., input step 0.2 V, ÷4 synthesis.

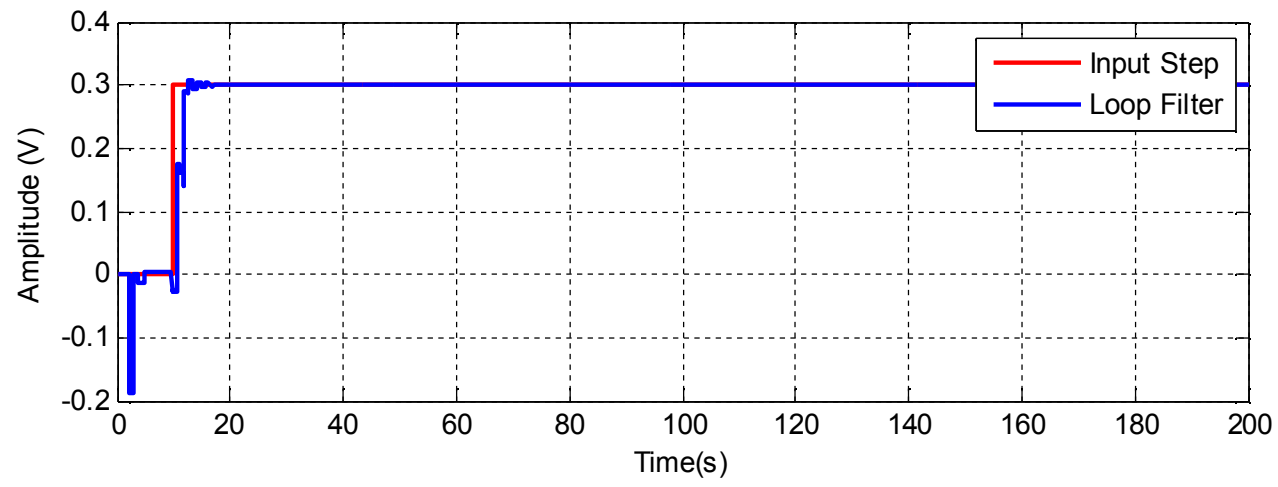

Fig. 10: LF O/P vs. input step of 0.3 V, ÷4.2 synthesis.

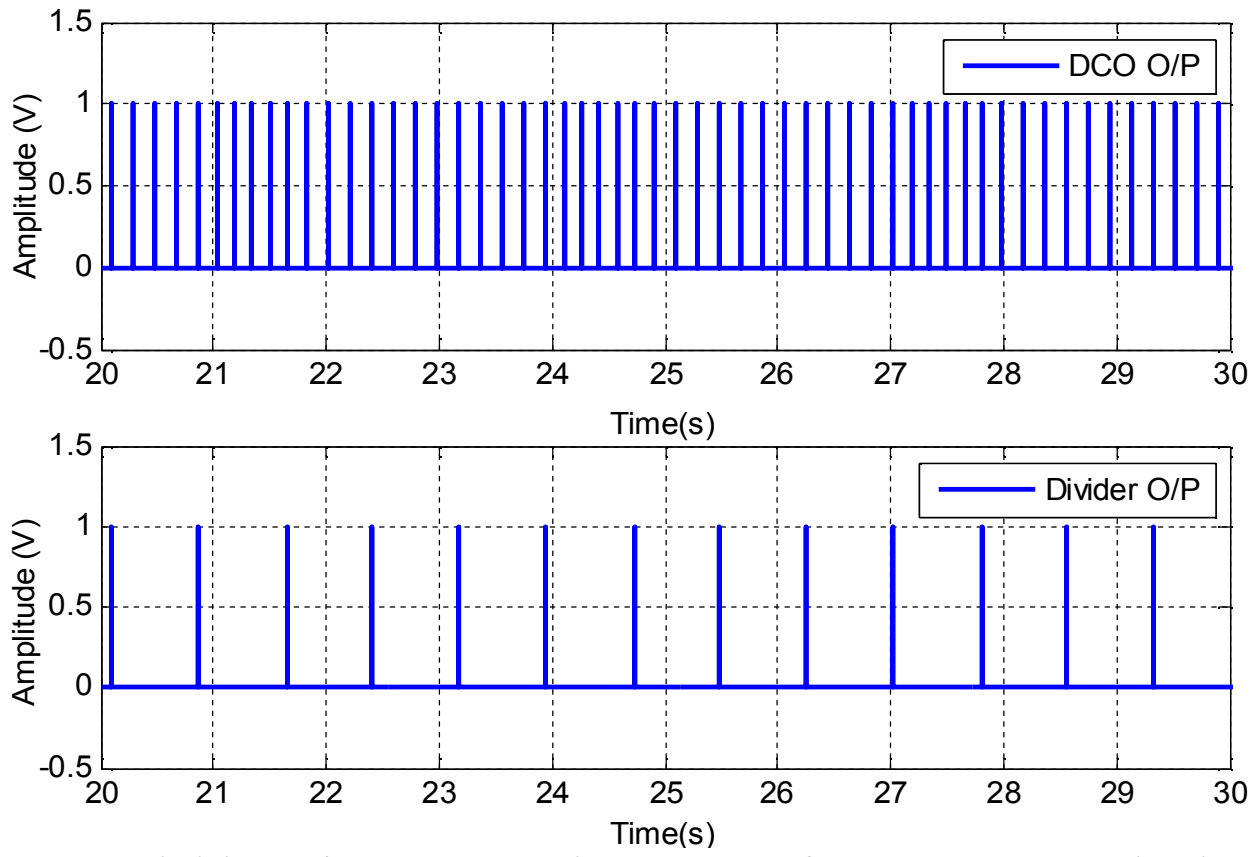

Fig. 11: Divider O/P. vs. DCO, input step of 0.3 V, ÷4.2 synthesis.

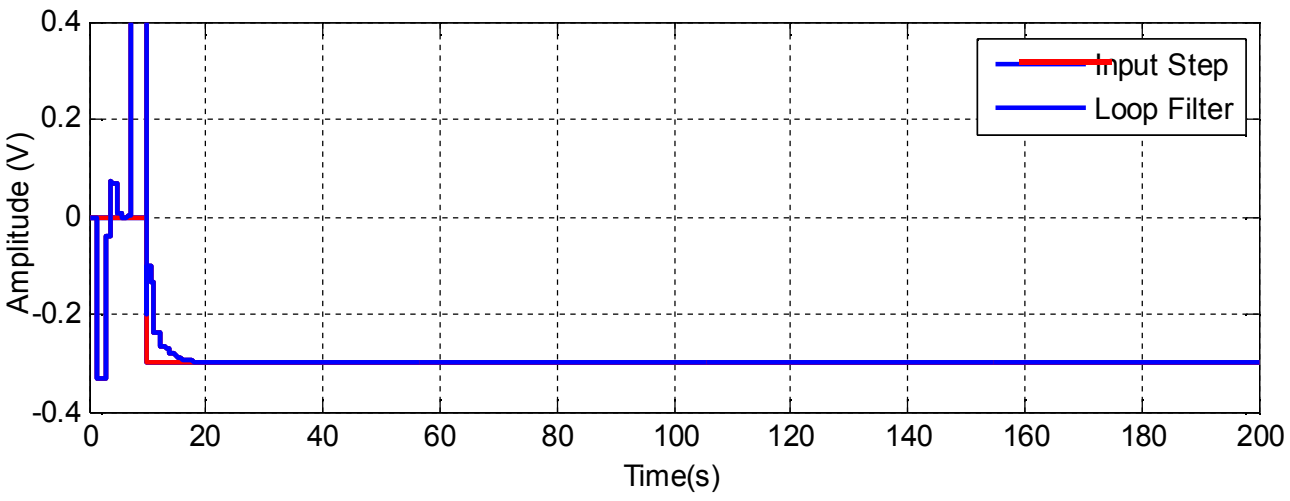

Fig. 12: LF O/P vs. input step of -0.3 volts, ÷3.125 synthesis.

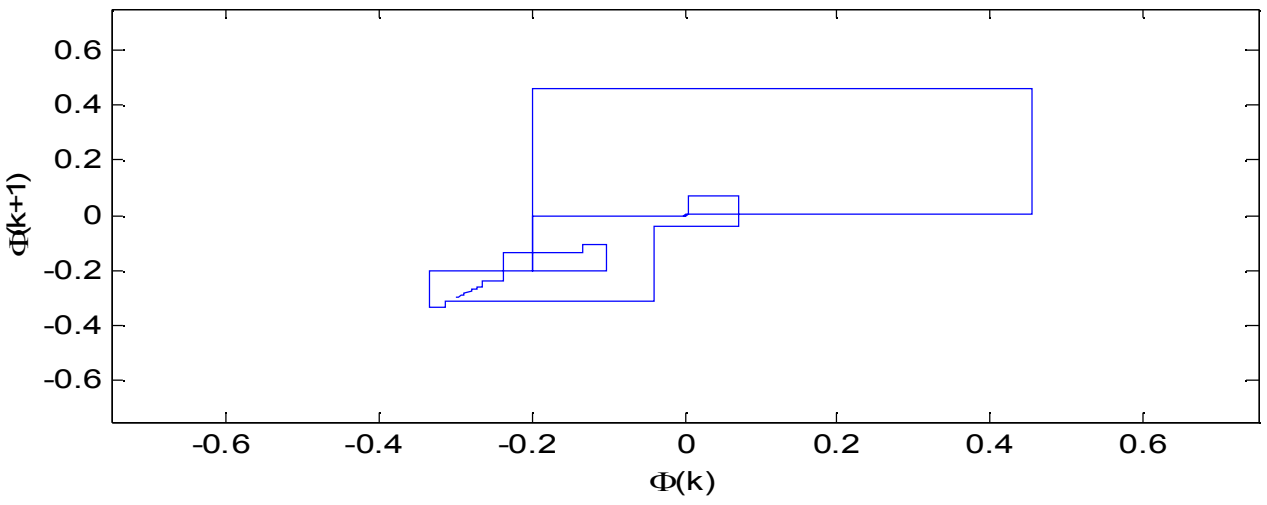

Fig. 13: Phase Plane for step of -0.3 volts, ÷3.125 synthesis

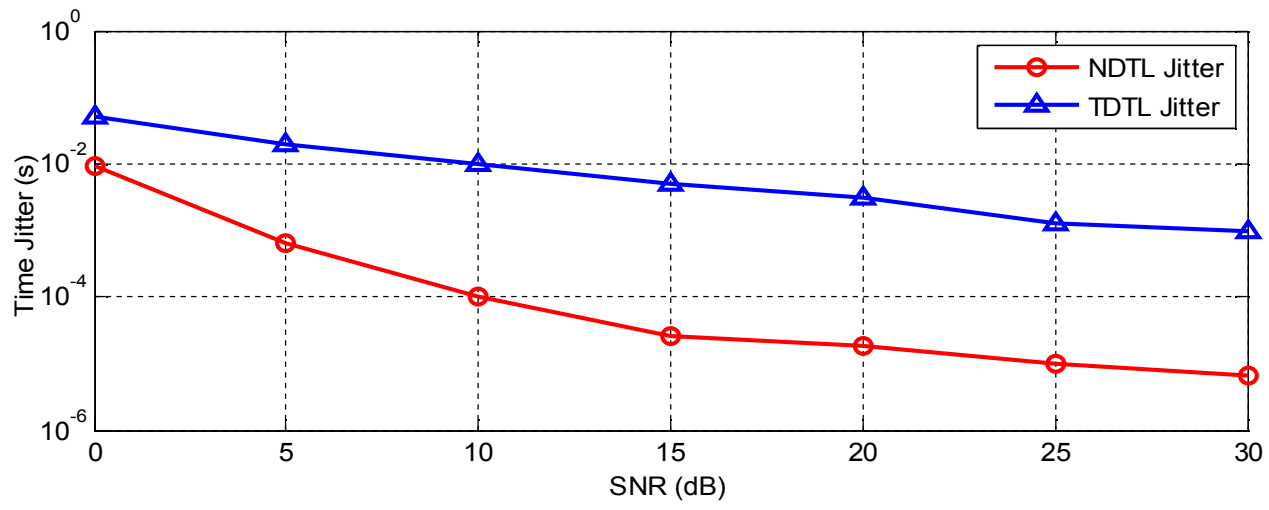

Figure 14: TDTL vs. NDTL jitter comparison for input step=0.3, *M*=3.125, VCO freq.=100 Hz, VCO sensitivity = 32 V/Hz, and division factor = 4.2.